\documentclass[preprint,12pt]{article}

\usepackage[pdftex]{graphicx}
\usepackage{ccaption}
\captiondelim {.  } 

\usepackage[centertags]{amsmath}
\usepackage{amsthm,amsfonts,amssymb}
\usepackage{indentfirst}
\usepackage[font={scriptsize}]{caption}

\newcommand{\bfM}{\mathbf{M}}
\newcommand{\bfm}{\mathbf{m}}
\newcommand{\bfu}{\mathbf{u}}
\newcommand{\bfv}{\mathbf{v}}
\newcommand{\bfo}{\mathbf{1}}
\newcommand{\bfD}{\mathbf{D}}
\newcommand{\bfQ}{\mathbf{Q}}

\newcommand{\bfp}{\mathbf{p}}

\newtheorem{theorem}{Theorem}[section]
\newtheorem{conj}{Conjecture}[section]

\title{Open Quasispecies Systems: New  Approach to Evolutionary Adaptation}

\date{\vspace{-5ex}}

\author{ 	
	{\bf Igor Samokhin} \\
	Lomonosov Moscow State University\\ Moscow 119992, Russia\\
	{\bf Tatiana Yakushkina}\\ National Research University Higher School of Economics\\
	Moscow 101000, Russia\\
	{\bf Alexander S. Bratus}\\
	Russian University of Transport\\Moscow 127994, Russia\\
	alexander.bratus@yandex.ru}

\begin{document}
	\maketitle

\begin{abstract}
Consider a mathematical model of evolutionary adaptation of fitness landscape and mutation matrix as a reaction to population changes. As a basis, we use an open quasispecies model, which is modified to include explicit death flow. We assume that evolutionary parameters of mutation and selection processes vary in a way to maximize the mean fitness of the system. From this standpoint, Fisher's theorem of natural selection is being rethought and discussed. Another assumption is that system dynamics has two significant timescales. According to our central hypothesis, major evolutionary transitions happen in the steady-state of the corresponding dynamical system, so the evolutionary time is much slower than the one of internal dynamics. For the specific cases of quasispecies systems, we show how our premises form the fitness landscape adaptation process. 
\end{abstract}

\section{Introduction}
\label{s1}
Adaptation is a fundamental property of all living systems. The very concept of species evolution implies the ability to respond to environmental changes. Many evolutionary models, like the Eigen quasispecies or Crow-Kimura models \cite{ei71,ei79,ki58} (an exhaustive review of this topic can be found in \cite{domingo2016}), consider populations of fixed size. However, this assumption can misrepresent the biological picture, and it is often crucial to take  growth and mortality properties into account \cite{bull08, kondr16, domingo2015, Park2010}. One of the examples, when we need a more accurate model, is bacterial population dynamics under medical treatment \cite{lev00}. In this scenario, death rates are inhomogeneous since the therapeutic effect is targeting specific pathogenic types. Another area of applications is cancer evolution \cite{kom14, kom15}. The recent studies \cite{mich14, han17} showed the development of chemotherapy-resistant cancer cells after a series of treatments. It is of significant interest to examine, at least for simplified cases, how such systems react to deliberate elimination of species. 

In this paper, we analyze replicator systems with explicit death rates and without constant population size condition. We call this class of systems  \textit{open replicator systems}.  In \cite{pavl12}, open replicator systems were first formalized to describe a spatially distributed population. Later, in \cite{bra19, yeg20}, open quasispecies systems were carefully studied. Here, we focus on dynamical properties of the system's fitness landscape. To move from static fitness landscape assumption,  we introduce fitness landscape adaptations and consider its  fluctuations. The question arises: how any adaptive changes are achieved in evolution? The central hypothesis of this study is that the specific time of the evolutionary adaptation of the system parameters is much slower than the time of the internal evolutionary process, which leads the system to its steady-state. Throughout the paper, we will call the first ``slow'' time the \textit{evolutionary time}. This assumption leads to the significant fact that the evolutionary changes of the system parameters happen in steady-states of the corresponding dynamical systems. The approach was previously used for hypercycles  \cite{drozh18} and later was applied to general replicator systems \cite{drozh19}.  In the current research, we address the question how open quasispecies systems react on directed mortality. 

\subsection{Open quasispecies model}
Consider a population with the distribution of types $\bfu(t) =$ $(u_1(t), \ldots,$ $u_n(t))$ changing over time $t \geq 0$. Here, $u_i(t)$ denotes a number of $i$-th type species, $i = 1,2, \ldots n$. For any particular moment, the replication process is defined by the set of fitness landscape coefficients $\bfm = (m_1, \ldots, m_n)$. We write it in a matrix form $\bfM = diag(\bfm) \in \mathbb{R}^{n\times n}.$ The death rates and mutation coefficients can be rewritten as well: 
$\bfD = diag(d_1, \ldots, d_n), $
$$\bfQ = \left\{q_{ij}: q_{ij}\geq 0, i\neq j, q_{ii}>0,\sum_{i=1}^nq_{ij}=1, \quad i,j=1,\ldots, n\right\}.$$ Here, $q_{ij}$ stands for the probability of having type $i$ as a result of replication of type $j$.   

To describe limitations of available resources in the system, we introduce a growth saturation function smooth $\phi(S)$: a non-negative function with  the domain  $S\in [0, +\infty)$, such that: 
$S\phi( S) = 0$ is a bounded function for $S\geq 0$. Without loss of generality, we suppose $\phi(S) = e^{-\gamma S}, S(t) = \sum_{i=1}^{n}u_i(t), \gamma = const>0.$

In this setting, an open quasispecies system can take the form of nonlinear differential equations: 
\begin{eqnarray}
\label{eq1}
&&\frac{d\bfu(t)}{ dt} = e^{ -\gamma S(t)} \bfQ_m \bfu(t) - \bfD\bfu(t), \quad \bfQ_m = \bfQ\bfM,\\
&&\bfu(0) = \bfu^0 > 0, \quad S(t) = \sum\limits_{ i = 1} ^ n u_i( t), \quad \gamma > 0.\nonumber
\end{eqnarray}
In \cite{yeg20}, it was shown that the system (\ref{eq1}) exhibits positive invariance in $\mathbb{R}^n_{+}$ and has a unique solution for $t\geq 0$ for all initial conditions $u^0\in \mathbb{R}^n_{+}.$
 
A similar logic applies to the Crow-Kimura system with the population distribution $\bfp$ over Hamming classes, which can be written as the open system: 
\begin{eqnarray}
\label{eq2}
&&\frac{d\bfp(t)}{dt} = e^{-\gamma S(t)}\left(\bfM+\mu {\bf G}\right)\bfp(t)- \bfD \bfp(t), \\
&&\bfp = (p_0, \ldots, p_N), \bfM = diag(m_0, \ldots, m_N), \bfD = diag(d_0, \ldots, d_N).\nonumber
\end{eqnarray}  
In this case, the mutation process is defined by the mutation rate parameter $\mu>0$ and the transition matrix:
$$
{\bf G} = \left(
\begin{array}{ccccccc}
-N & 1 & 0 & 0& \ldots& 0 & 0\\
N & -N & 2 & 0 & \ldots& 0 & 0\\
0 & N-1 & -N & 0 &\ldots & 0 & 0\\
0 & 0 & N-2& -N &\ldots & 0 & 0\\
\ldots & \ldots & \ldots& \ldots &\ldots & \ldots & \ldots\\
0 & 0 & 0 & 0 &\ldots&N-1 & 0\\
0 & 0 & 0 & 0 & \ldots & -N & N\\
0 & 0 & 0 & 0 & \ldots & 1 &-N\\
\end{array}\right).
$$

For the quasispecies system (\ref{eq1}), the mean fitness is defined as follows \cite{yeg20}: 	
\begin{eqnarray}
\label{eq3}
f(t) = \displaystyle\left\{
\begin{array}{lc}
 0, & \sum_{i=1}^{n}u_i=0,\\
\displaystyle\frac{\sum\limits_{i = 1}^n m_iu_i(t)}{\sum\limits_{i = 1}^n d_iu_i(t)}, & \sum_{i=1}^{n}u_i>0.
\end{array}
\right.
\end{eqnarray}
The numerator of this fitness function coincides with the mean fitness in the classical quasispecies model. The denominator (\ref{eq3}) denotes the total population loss due to the death rate in the system (\ref{eq1}). One can note that an analogous expression for the mean fitness is valid for the open Crow-Kimura system. 

In the following discussion, we consider the steady-state $\bar{\bfu}$ of the system (\ref{eq1}), which can be described by the equation: 
\begin{equation}\label{eq4}
\bfD^{-1}\bfQ_m\bar{\bfu} = e^{\gamma S(\bar{\bfu})}\bar{\bfu}, \quad\gamma >0.
\end{equation}
The latter expression, in contrast to a similar one for the standard Eigen model, does not represent an eigenvalue problem for $\bfD^{-1}\bfQ_m$ in a traditional sense.

The components of the steady-state $\bar{\bfu} = \left(\bar{u}_1, \ldots, \bar{u}_n\right)$ are such that: 
$$
\bar{u}_i = \lim\limits_{T\rightarrow\infty}\frac{1}{T}\int_{0}^{T}u_i(t)dt, \quad i=1,\ldots,n.
$$
In \cite{yeg20}, it was shown that (\ref{eq4}) can have a simple eigenvalue $\lambda^*$ with maximal absolute value for irreducible matrices $\bfQ_m$. For such eigenvalue, there is a leading  eigenvector $\bfu^* \geq 0$ such as 
\begin{eqnarray}\label{e5}
S(\bar{\bfu}) = \gamma^{-1}\ln \lambda^*,\quad
\lambda^* = \frac{\sum\limits_{i = 1}^n m_i\bar{u}_i}{\sum\limits_{i = 1}^n d_i\bar{u}_i}.
\end{eqnarray} 
The latter expression, according to (\ref{eq3}), defines the mean fitness value of the system (\ref{eq1}). A similar result can be derived for the Crow-Kimura setting. 

\subsection{Assumptions for fitness landscape optimization}\label{s12}
In this paper, following the previous studies on different classes of replicator systems \cite{drozh18, drozh19}, we propose the two main assumptions.
\begin{itemize}
	\item  For open quasispecies systems, evolutionary adaptation of fitness landscape and mutation matrix to low death rates variation satisfies Fisher's fundamental theorem of natural selection \cite{fi30}. This means that it results in mean fitness maximization.  
	\item The time of evolutionary changes is much slower than the one of the internal system dynamics. That is, there are two timescales: first describes the dynamics of the system with particular parameters up to the steady-state; second --- the process of small evolutionary changes. 
\end{itemize}
The latter means that adaptation of the system's parameters happens in a series of steady-states. The evolutionary timescale $\tau$ defines the parameter of such adaptations.  
In other words, the problem of evolutionary adaptation of the replicator system to environmental changes (in this particular case expressed in death rate variations)  leads to choosing such system parameters as functions of $\tau$, that they maximize the mean fitness (\ref{e5}). As it was shown in \cite{drozh19}, it can be interpreted as searching for the combination of parameters that ensure the eigenvalue maximum (\ref{e5}). This class of problems is  widespread in different areas of physics and mechanics, where leading eigenvalue defines first normal mode of oscillation or stability loss rate in dynamical systems. 

In biological literature, there is still an ongoing discussion on extremum principals in evolution \cite{ao05}. Different interpretations of Fishers's theorem of natural selection \cite{ew89, less97} ans Wright's concept of adaptive fitness landscape \cite{ao09} are being examined and applied. Theoretical results depend heavily on mathematical formalization of the replication process. Many of the results lie in the field of evolutionary game theory \cite{hz88, ho02}. For classical replicator equations with symmetric interaction matrix, the mean fitness is proven to be a monotonically increasing function \cite{Cress92}.  In a more general case of replicator systems, the mean fitness does not have to be monotonic and can decrease locally before reaching the maximum state, which is not necessarily the steady-state value. In \cite{bra18}, the authors obtained necessary and sufficient conditions for the maximum value of the mean fitness and its value in the steady-state to coincide for a general case of replicator equations.   

\section{Evolutionary adaptation}
Consider an open quasispecies system (\ref{eq1}) under the assumptions on its adaptation  made above. 
Let the death rates $\bfD = diag(d_1, \ldots, d_n)$ be fixed over evolutionary time $\tau$, $d_i >0, i=1,\ldots, n$. We assume that $\bfM = \bfM(\tau)$ and $\bfQ = \bfQ(\tau)$ are smooth functions with slow growths with respect to $\tau$. Moreover, the landscape variations are limited by a restriction on resources $K>0=const$: 
$$\bfM(\tau) = diag\left(m_1(\tau),\ldots, m_n(\tau)\right),\quad m_i(\tau) \geq 0, i=1,\ldots, n.$$
\begin{equation}\label{eq6}
\bfM_K(\tau) = \left\{\bfM(\tau): \sum_{i=1}^{n}m_i(\tau)\leq K\right\}.
\end{equation}
For mutation matrix, we denote a set:
$$
{R(\tau)} = \left\{q_{ij}(\tau):  q_{ij} \geq 0, i\neq j, q_{ii} \geq \delta_{ij} > 0\sum^{n}_{i=1}q_{ij}(\tau) = 1, \right\}. 
$$
In terms of the vector-function  $\bar{\bfu}(\tau) \in \mathbb{R}^n_{+}$ in a steady-state, we get an equation in evolutionary time $\tau>0$:
\begin{equation}\label{eq7}
\bfD^{-1}\bfQ_m(\tau)\bar{\bfu}(\tau) =  e^{\gamma S(\bar{\bfu}(\tau))}\bar{\bfu}(\tau).
\end{equation}
Here, $\bfQ_m(\tau) = \bfQ(\tau)\bfM(\tau),$ $\bar{\bfu}(\tau) = \left(\bar{u}_1(\tau),\ldots, \bar{u}_n(\tau)\right),$ $S(\bar{\bfu}) = \sum_{i=1}^{n}\bar{u}_i$
Taking into account the latter equality (\ref{eq7}) and the following property:
$$
\sum_{i=1}^{n}\left(\bfQ_m\bar{\bfu}\right)_i = \sum_{i,j=1}^{n}q_{ij}m_j\bar{ u}_j = \sum_{i=1}^{n}m_i\bar{ u}_i,
$$ 
we get that:
\begin{equation}\label{eq8}
 e^{\gamma S(\bar{\bfu})} = \frac{\sum\limits_{i = 1}^n m_i\bar{u}_i}{\sum\limits_{i = 1}^n d_i\bar{u}_i}
\end{equation}
Taking into account the definition of the mean fitness in open quasispecies system (\ref{eq3}) and its value in a steady-state (\ref{e5}), we can interpret the latter expression (\ref{eq8}) as the mean fitness in a steady-state. This means that 
\begin{equation}\label{eq9}
\bar{ f}(\tau) = e^{\gamma S(\bar{\bfu})}, S(\bar{\bfu}) = \frac{1}{\gamma}\ln\bar{ f}(\tau)
\end{equation} 

Consider a problem of mean fitness maximization $\bar{f}(\tau)$ over the set of possible fitness landscapes $\bfM_K(\tau)$ and mutation matrices $R(\tau)$, where the death rate matrix $\bfD$ remains constant. This maximization problem is examined in the evolutionary timescale dynamics $\tau \geq 0$. As an initial condition at $\tau = 0$, we take a fitness landscape with parameters $\bfM_K(0)$ and $R(0)$. The rest of the optimization process can be described as a sequence of steps $\tau$: at each step, fitness landscape parameters are chosen from $\bfM_K(\tau)$ and $R(\tau)$ in order to maximize the mean fitness $\bar{f}(\tau)$.

It is worth pointing out that for any fixed value $\tau>0$, we can reconstruct the dynamics of the system and find the distribution of the population $\bfu(t,\tau)$. Since the system is permanent, we can obtain the trajectories by solving the following equations: 
\begin{eqnarray}\label{eq10}
&&\frac{d\bfu(t,\tau)}{dt} = e^{-\gamma S(t, \tau)} \bfQ_m \bfu(t,\tau) - \bfD\bfu(t,\tau),\nonumber\\ 
&&\bfQ_m(\tau) = \bfQ(\tau)\bfM(\tau), \bfQ(\tau)\in R(\tau), \bfM(\tau)\in \bfM_K(\tau),\\
&&u(0,\tau) = u^0(\tau) > 0, \quad S(t, \tau) = \sum\limits_{ i = 1} ^ n u_i(t,\tau), \quad \gamma > 0.\nonumber
\end{eqnarray}

In the Crow-Kimura case, we have similar expressions to (\ref{eq7}, \ref{eq9}):
\begin{eqnarray}\label{eq11}
\bfD^{-1}\left(\bfM(\tau)+\mu {\bf G}\right)\bar{\bfp} = e^{\gamma S(\bar{\bfp})}\bar{\bfp},\\
S(\bar{\bfp}) = \gamma^{-1}\ln\lambda^*(\tau). \nonumber
\end{eqnarray}

Does the fitness function $\bar{f}(\tau)$ (\ref{eq9}) reach a maximum over the considered set? Applying known result for an essentially non-negative matrix \cite{cohen82}, we obtain that for $\tau>0$ the mean fitness of quasispecies system is a convex function with respect to elements of the fitness matrix $M(\tau)$. Hence, there is a global maximum of the mean fitness function $\bar{ f}(\tau)$, which is reached at a peak of $\bfM_K(\tau)$: $m_{peak} = K,$ and  $m_i=0$ for all other species. 

Similar to the case (\ref{eq4}), the equation (\ref{eq7}) is not a classical eigenvalue problem for $\bfD^{-1}\bfQ_m$. Indeed, a substitution $\xi\bar{\bfu}, \xi \in \mathbb{R} $ for $\bar{ u}$ does not lead to a homogeneous equation for $\xi$. For a fixed evolutionary time moment $\tau,$ consider the following eigenvalue problem: 
\begin{equation}\label{eq12}
\bfD^{-1}\bfQ_m(\tau) \bar{\mathbf{w}}(\tau) = \lambda(\tau)\bar{\mathbf{w}}(\tau).
\end{equation}
Matrix  $\bfD ^ {-1} \bfQ_m$ is positive, hence, Frobenius-Perron theorem suggests $\lambda > 0, \quad \bar{ u} > 0$. However, (\ref{eq9}) gives $\bar{S} \geq 0$. Let us show that  the latter holds if the diagonal elements of the matrix  $D$ are small enough. It is known in the literature  \cite{Bellman_1976}, that for positive matrices $A_1$ and $A_2$ such that  $A_1 \leq A_2$, one has $\lambda( A_1) \leq \lambda( A_2)$. When diagonal elements of $\bfD$ are decreasing, positive eigenvalues of the matrix $\bfD^ {-1} \bfQ_m$ are increasing monotonically. Hence, there is a non-empty set of diagonal matrices  for which $\lambda( \bfD^ {-1} \bfQ_m) \geq 1$, e.g.,  $\bar{ S} \geq 0$. We assume that the pre-defined matrix $\bfD$ is chosen to satisfy this condition for $\tau \geq 0.$

\begin{theorem}\label{thm21}
	Let $\eta >1$ and solution of the problem (\ref{eq7}) satisfy the condition: 
	\begin{equation}\label{eq13}
	\sum_{i=1}^{n}\bar{u}_i(\tau) = S(\bar{\bfu}(\tau)) = \frac{1}{\gamma}\ln\eta, \gamma >0
	\end{equation}
	for a fixed  $\tau \geq 0.$ Then, a unique non-trivial solution $\bar{\bfu}(\tau)\in\mathbb{R}^n_{+}$ to the equation (\ref{eq7}) over the set (\ref{eq13}) exists if and only if $\lambda(\tau) = \eta >1.$ 
\end{theorem}
Here, $\lambda(\tau)$ is a maximal positive eigenvalue (\ref{eq12}) at fixed $\tau\geq 0$. 
For the proof of this theorem, see the \ref{app1}.

For open quasispecies systems in the form (\ref{eq10}), the following theorem takes place.
\begin{theorem}\label{thm22}
	Let $\bfD = diag(d_1, \ldots, d_n)$, $d_i>0, i=1,\ldots, n,$ be a constant matrix, such as 
	$\lambda^*(\bfD^{-1}\bfQ_m)>1$ holds for any $\bfM\in \bfM_K(\tau)$ and $\bfQ(\tau)\in R(\tau), \tau\geq 0.$ Then, the solution to (\ref{eq7}) belongs to a convex set for each $\tau\geq 0$:
	\begin{equation}\label{eq14}
	U_\tau = \left\{\bar{\bfu}(\tau)\in \mathbb{R}^n_{+}, S(\bar{\bfu}(\tau)) = 
	\sum_{i=1}^{n}u_i(\tau)\leq \hat{ S} = \gamma^{-1}\ln\frac{K}{\check{d}}\right\},
	\end{equation}
	where, $\check{d} = \min\left\{d_1, \ldots, d_n\right\}$.	 
\end{theorem}
It is easy to show, that from (\ref{eq9}) it follows:
\begin{equation}\label{eq15}
S(\bar{\bfu}) = \gamma^{-1}\ln\frac{(m,\bar{\bfu}(\tau))}{(\bfD\bar{\bfu}(\tau), \bfo)} \leq \gamma^{-1}
\ln\frac{K}{\check{d}} = \hat{ S}, \bfo = (1,\ldots, 1).
\end{equation}
\begin{conj}\label{con21}
	If the conditions of Theorem \ref{thm22} apply, then for each $\tau\geq 0$ $\ln\lambda(\bar{ \bfu}(\tau)) = \ln\bar{f}(\bar{\bfu}(\tau))$ is a linear functional over the set $U_\tau.$
\end{conj}
Here, $\lambda(\bar{\bfu}(\tau)) = \bar{f}(\bar{\bfu}(\tau))$ is a dominant eigenvalue of the problem (\ref{eq7}).  Indeed, if $\bar{\bfu}$ is a solution to (\ref{eq7}) , then (\ref{eq9}) holds true. Hence, 
$S(\xi\bar{\bfu}) = \xi S(\bar{\bfu}) = \xi\gamma^{-1} \ln\lambda^*(\bar{\bfu})$. At the same time, $S(\xi\bar{\bfu}) = \gamma^{-1}\ln\lambda^*(\xi\bar{\bfu}).$ 

\begin{conj}\label{con22}
	If the conditions of Theorem \ref{thm22}  apply , then for each $\tau\geq 0$ over the convex set $U_\tau$ there exists a unique maximum value $\ln\bar{f}(\bar{\bfu}(\tau)) = \gamma S(\bar{\bfu}),$ $\bar{\bfu} \in U_\tau$.
\end{conj}

\section{Fitness variation and  necessary condition for extremum}

Let us construct the conditions for the mean fitness function $\bar{f}(\tau)$ calculation. 
Taking into account the assumptions discussed above for the mean fitness maximization problem, we can apply the well-known results \cite{kato} for one-parameter spectrum perturbation for the matrix $\bfD^{-1} \bfQ_m(\tau), \tau>0$. This means that the corresponding eigenvalue and eigenvector of this matrix can be decomposed into a series with a small perturbation parameter. 
We use the notation $\delta \bfM(\tau)$, $\delta \bfQ(\tau)$, $\delta \bar{\bfu}(\tau)$, and $\delta\lambda^*(\tau)$ for principal linear part in the increment of fitness landscape parameters  $\bfM(\tau)$, $\bfQ(\tau)$, vector-function $\bar{\bfu}(\tau)$ and eigenvalue $\lambda(\tau)$ in evolutionary time increment $\tau\rightarrow \tau+\delta\tau, \delta\tau>0.$ 

\begin{eqnarray}\label{eq16}
\bfM(\tau + \delta\tau) = \bfM(\tau) + \delta \bfM\delta\tau + o(\delta\tau),
\bfQ(\tau + \delta\tau) = \bfQ(\tau) + \delta \bfQ\delta\tau + o(\delta\tau),\\
\bar{\bfu}(\tau + \delta\tau) = \bar{\bfu}(\tau) + \delta \bar{u}\delta\tau + o(\delta\tau),
\lambda^*(\tau + \delta\tau) = \lambda^*(\tau) + \delta \lambda^*\delta\tau + o(\delta\tau).\nonumber
\end{eqnarray}
Since $\bfM(\tau) \in \bfM_K(\tau)$ and $\bfQ(\tau)\in R(\tau)$, then for the elements of $\delta \bfM$ and $\delta \bfQ$ it is necessary to satisfy the conditions: 
\begin{equation}\label{eq17}
\sum_{i=1}^{n}\delta m_i(\tau) =0, \delta m_i \geq 0 \textrm{ for } m_i=0, \sum_{i=j}^{n}\delta q_{ij}(\tau)=0, \delta q_{ii}\geq 0.
\end{equation}
If $q_{ij} = 0$, then $\delta q_{ij}\geq 0, i\neq j, i,j=1,\ldots, n.$ Substituting decomposition (\ref{eq16}) into (\ref{eq7}) and keeping only linear part for $\delta \tau$, we get:  
\begin{equation}\label{eq18}
\bfD^{-1}\left(\delta \bfQ\bfM\bar{\bfu} + \bfQ\delta \bfM\bar{\bfu} + \bfQ_m\delta\bar{\bfu}\right) = \delta \lambda^*\bar{\bfu} + \lambda^*\delta\bar{\bfu}.
\end{equation}
Introduce an adjoint problem to (\ref{eq7}) eigenvalue problem, having $\lambda^*(\tau)\in \mathbb{R}_{+}$:
\begin{equation}\label{eq19}
(\bfD^{-1} \bfQ_m(\tau))^T \bar{\bfv}(\tau) = \lambda^*\bar{\bfv}(\tau).
\end{equation}
Without any loss of generality, we assume the vector $\bar{\bfv}$ being normalized:
\begin{equation}\label{eq20}
(\bar{\bfu}, \bar{\bfv}) =1, 
\end{equation}
here and throughout the paper we use brackets for scalar product. 

Multiplying the equation (\ref{eq18}) by eigenvector of the problem (\ref{eq19}) and using (\ref{eq20}), together with the expression:
\begin{equation}\label{eq21}
\bfD^{-1}(\delta \bfQ_m) = \bfD^{-1}\left(\delta \bfQ\bfM+\bfQ\delta \bfM\right),
\end{equation}
we get equation for $\delta\lambda^*(\tau)$ and the mean fitness value variation:
\begin{equation}
\label{eq22}
\delta \bar{f}(\tau) = \delta \lambda^*(\tau) = \left(\bfD^{-1}\delta\bfQ_m(\tau)\bar{ \bfu}(\tau),\bar{\bfv}(\tau)\right).
\end{equation}

From these derivations, we show that the original problem of mean fitness maximization in a steady-state transformed into a linear programming problem. That is, maximization of $\lambda^*(\tau)$ or $\bar{ f}(\tau)$ over the interval $(\tau, \tau +\delta\tau]$ takes the form:
\begin{equation}\label{eq23}
\delta \bar{f}(\tau) = \left( \bfD^{-1}\delta \bfQ_m(\tau)\bar{\bfu}(\tau), \bar{ \bfv}(\tau)\right)\rightarrow \max,
\end{equation}
with restrictions (\ref{eq17}). Necessary condition for extremum requires the left-side of the latter expression (\ref{eq22}) to be zero for all $\delta m_i, \delta q_{ij}$ under the condition (\ref{eq17}). If we include the higher-order terms $(\delta\tau)^2,$ then we obtain the exact form of extremum (see \ref{app2}).

\subsection{Fitness landscape variation}
Necessary extremum condition would be significantly simpler for the case when mutations do not change over evolutionary time $\tau$. In this particular scenario, $\delta \bfQ = 0,$ and the necessary  condition for extremum transforms into:
\begin{equation}\label{eq24}
\delta\bar{f}(\tau) = \left(\bfD^{-1}\delta\bfM(\tau)\bar{\bfu}(\tau), \bar{\bfv}(\tau)\right) = 0,
\end{equation}
over a set 
\begin{equation}\label{eq25}
\sum_{i=1}^{n}\delta m_i(\tau) =\left(\delta \bfm(\tau), \bfo\right)=0, \quad \delta m_i\geq 0 \textrm{ for } m_i=0.
\end{equation}
We denote $\delta \bfm(\tau) = \left(\delta m_1, \ldots, \delta m_n\right), \delta m_i= m'_i(\tau), \bfo = (1,\ldots,1).$ 
The equality (\ref{eq24}) can be rewritten as:
\begin{equation}\label{eq26}
\left(\delta \bfm, \bar{l}\right) = 0,\quad \bar{l} = diag\left(\bar{\bfu}\bfD^{-1}\bfQ^T\bar{\bfv}\right).
\end{equation}
For $(\delta\bfm, \bfo)=0$, the latter condition means that $\bar{l}$ has to be collinear with $\bfo:$ $\bar{l} = c\bfo, c = const.$
This condition works for a local extremum for quasispecies systems. 

For numerical simulations with an iteration step $\varepsilon$, we use the linear programming problem in the form: 
\begin{eqnarray}\label{eq27}
&&\left(\delta \bfm, \bar{l}\right)\rightarrow \max\\
&&\sum\limits_{ j = 1} ^ n \delta m_j = 0, \quad  \max( - \varepsilon \bfo, - \bfm) \leq \delta \bfm \leq \varepsilon \bfo\nonumber.
\end{eqnarray}

In the case of Crow-Kimura equations, the same condition is necessary and sufficient due to convexity of the problem. 
As for (\ref{eq24}), the analogous derivation gives:
\begin{equation}\label{eq28}
\delta\bar{f}(\tau) = \left(\bfD^{-1}\delta\bfM(\tau)\bar{\bfp}(\tau), \bar{\bfv}(\tau)\right).
\end{equation}
Similar to (\ref{eq21}),  $\bar{ \bfv}$ is the solution to adjoint problem:
\begin{equation}\label{eq29}
\bfD^{-1}\left(\bfM+\mu{\bf G}\right)^T\bar{\bfv} = \lambda^*\bar{\bfv}.
\end{equation}

\subsection{Mutation matrix variation}
Consider a case when the fitness landscape and death rates are constant (
$\delta\bfM = 0$  and $\bfD=const$). Let us examine variation of the mutation  matrix $\bfQ + \delta\bfQ$ at the steady-state $\bar{\bfu}$, where $|\delta q_{ij}|\leq \varepsilon.$
\begin{equation}\label{eq30}
\delta\bar{f} = \sum\limits_{i, j = 1} ^ { n} ( \delta q_{ i, j}, m_j \bar{u}_j \bar{ v}_i).
\end{equation}

For numerical simulations in this case, the linear programming problem takes the form: 
\begin{eqnarray}\label{eq31}
&&\delta\bar{ f}\rightarrow \max\\
&&\sum\limits_{ j = 1} ^ n \delta q_{ij} = 0, \quad  \max( - \varepsilon \mathbf{E}, - \bfQ + \min(q_{ij})\mathbf{E}) \leq \delta \bfQ \leq \varepsilon \mathbf{E}.\nonumber
\end{eqnarray}

\section{Numerical simulations}
The expressions obtained above and the previous discussion show that the mean fitness $\bar{f}(\tau)$ maximization problem  transforms into an  linear programming iteration process (\ref{eq23}). The restrictions for this optimization are described   
by (\ref{eq17}).  Let $\varepsilon >0$ denote a step of such optimization process in evolutionary timescale: $\delta \tau = \varepsilon >0$.  Applying the hypothesis of slow adaptation pace for the  matrices $\bfM$ and $\bfQ$ over evolutionary time $\tau$ and taking small enough values $\varepsilon>0$:
\begin{eqnarray}\label{eq32}
|\delta m_i(\tau)| = |m_i'(\tau)|\leq m_0\varepsilon, |m_i''(\tau)|\leq m_1\varepsilon,\nonumber\\
|\delta q_{ij}| = |q_{ij}'(\tau)|\leq q_0\varepsilon, |q_{ij}''(\tau)|\leq q_1\varepsilon,\\	
|\delta \bar{f}(\tau)| = |\bar{f}_i'(\tau)|\leq f_0\varepsilon, |f_i''(\tau)|\leq f_1\varepsilon,	 \nonumber
\end{eqnarray}
where $m_0, m_1, q_0, q_1, f_0, f_1$ are constants with positive values.
Hence, all the equalities in decomposition (\ref{eq16}) hold with accuracy $\varepsilon^3$. 
If the number of steps in the iteration process is of order $\varepsilon^{-1}$, then the total error will have the order $\varepsilon^2$.

Let us consider a numerical scheme for the constructed optimization process with $n=2^4.$ The elements of mutation matrix are calculated according to the formula:
\begin{equation}\label{eq33}
q_{ij} = p^{4-\kappa_{ij}}(1-p)^{\kappa_{ij}}, i,j = 1,2,\ldots, 16.
\end{equation}
Here, $\kappa_{ij}$ is a Hamming distance between types $i$ and $j$, where each type is encoded by binary strains. For illustrations, we take the probability of errorless replications as $p=0.9.$ The expression (\ref{eq33}) is applied for a constant environment. During the adaptation process, every new structure of the mutation matrix at each step is chosen from the set $R(\tau)$.

\subsection{Numerical example 1}\label{ex1}
\begin{figure}[h!]
	\centering
	\includegraphics[width = 0.6\textwidth]{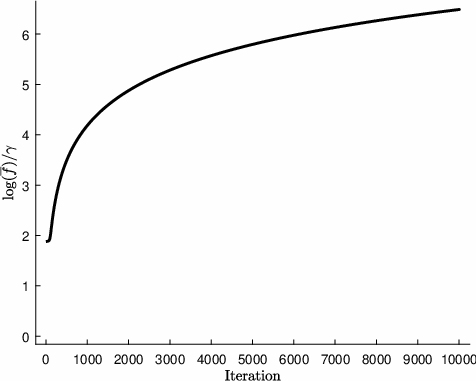}
	\caption{Example 1 (changing $\bfM$ and $\bfQ$): The mean fitness value changing over evolutionary time, which is represented by the number of iterations}
	\label{fig1}
\end{figure}

Consider the following set of parameters.
\begin{itemize}
	\item The initial distribution of fitness is concentrated at the first type: $\bfm_0: m_{01} = 1, m_{0i} =0, i=2,\ldots, 16$. Hence, the value $K$ in the restriction on available resources (\ref{eq6}) is $1$.
	\item Death rates:  $D = diag$(0.1, 0.001, 0.001, 0.00051, 0.001, 0.00051, 0.00051, 0.00034, 0.001, 0.00051, 0.00051, 0.00034, 0.00051, 0.00034, 0.00034, 0.00026).
	\item The first steady-state (before the evolutionary adaptation started): $\bar{ u}(0) =$(0.0304, 0.3386, 0.3386, 0.0742, 0.3386, 0.0742, 0.0742, 0.0122, 0.3386, 0.0742, 0.0742, 0.0122, 0.0742, 0.0122, 0.0122, 0.0018) with the total population size $S = 1.881.$
	\item Step for evolutionary timescale iteration: $\varepsilon = 0.0001$.
\end{itemize}
The ``length'' of the evolution is defined by the number of iterations. After $10^5$ iterations of the evolutionary adaptation process, we obtain that the fitness landscape fully changed. As a result, the fitness value increases steadily  over the evolutionary time, even though the growth rate reduces, as shown in Fig.~\ref{fig1}.The structure of the population changed and the second type dominated with $m_2=1, m_{i\neq 2} =0$. 
The end-state of this process is characterized by $\bar{u}_{end}=$ (0.0040, 0.0441, 3.5684, 0.7824, 0.0441, 0.0097, 0.7824, 0.1290, 0.0441, 0.0097, 0.7824, 0.1290, 0.0097, 0.0016, 0.1290, 0.0189) with $S(\bar{ u}_{end}) = 6.488$.	Thus, the mean fitness increased by  3.4 times approximately. 

\begin{figure}[h!]
	\begin{center}
		\includegraphics[width = 0.6\textwidth]{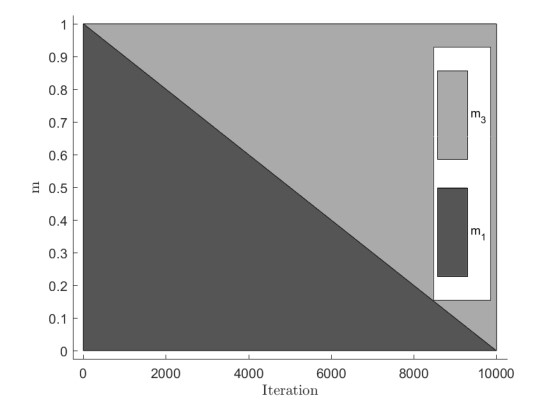}
		\caption{Example 1 (changing $\bfM$ and $\bfQ$): Dynamics of the fitness landscape parameters over evolutionary time: fitness matrix values $m_i$ in steady-states with respect to the number iterations}
		\label{fig2}
	\end{center}
\end{figure}

\begin{figure}[h!]
	\begin{minipage}[h]{0.5\linewidth}
			\center{\includegraphics[width=1\linewidth]{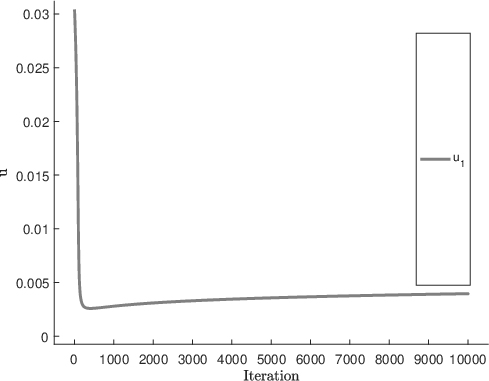} \\ a)}
	\end{minipage}
		\hfill
	\begin{minipage}[h]{0.5\linewidth}
			\center{\includegraphics[width=1\linewidth]{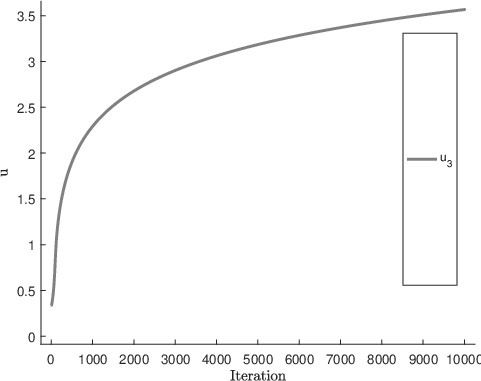}\\ b)}
	\end{minipage}
	\caption{Example 1 (changing $\bfM$ and $\bfQ$): Dependence of the steady-state distribution component a) $\bar{ u}_1$  and b) $\bar{ u}_2$ on the iteration number in the evolutionary timescale}
	\label{fig3}
\end{figure}

The dynamics of the fitness landscape components is shown in Fig.~\ref{fig2}. In Fig.~\ref{fig3}, we show how the rates of 1st and 3d species change over this process: the fraction of the first one, dominant in the beginning, drops down, while the latter prevails.  

\subsection{Numerical example 2}\label{ex2}
Let us move on to the case, when the fitness landscape remains constant while the mutation parameters vary over the evolutionary time. We take the same fitness parameters, death rates, iteration step, and initial distribution as in the previous example. We assume that the diagonal elements of the mutation matrix $q_{ii}\geq 0.6, \forall i.$ In this case, the mutation matrix fully changed after $3433$ iterations. 

\begin{figure}[h!]
	\centering
	\includegraphics[width = 0.6\textwidth]{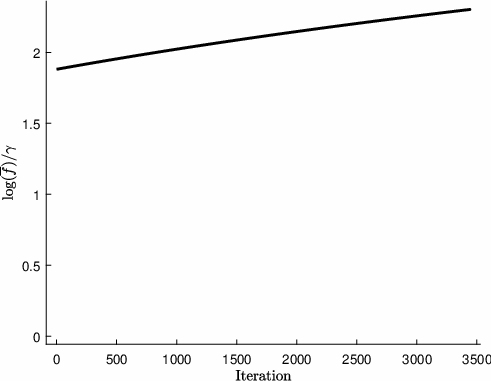}
	\caption{Example 2 (changing $\bfQ$): The mean fitness value changing over evolutionary time, which is represented by the number of iterations}
	\label{fig4}
\end{figure}

\begin{figure}[h!]
	\begin{minipage}[h]{0.5\linewidth}
		\center{\includegraphics[width=1\linewidth]{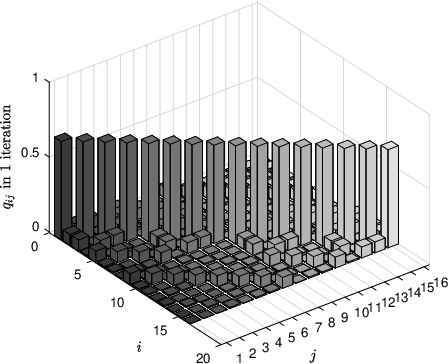} \\ a)}
	\end{minipage}
	\hfill
	\begin{minipage}[h]{0.5\linewidth}
		\center{\includegraphics[width=1\linewidth]{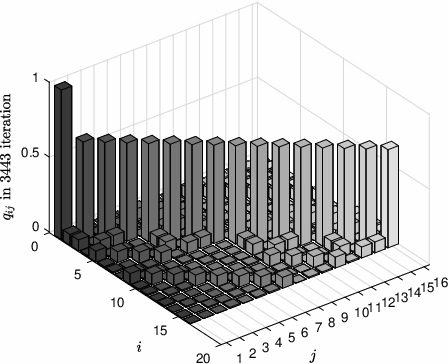}\\ b)}
	\end{minipage}
	\caption{Example 2 (changing $\bfQ$): The mutation matrix coefficients at a) the first and b)  3443th iterations of the evolutionary process}
	\label{fig5}
\end{figure}

In this case, the fitness value was increasing as well (Fig. \ref{fig4}) along the evolutionary timescale. Changes in the mutator matrix can be seen in Fig. \ref{fig5}.
The final state of this process is the vector $\bar{u}_{end}=$ (0.0563, 0.4110, 0.4110, 0.0901, 0.4110, 0.0901, 0.0901, 0.0149, 0.4110, 0.0901, 0.0901, 0.0149, 0.0901, 0.0149, 0.0149, 0.0022) with $S(\bar{ u}_{end}) = 2.303$.	Here, the total growth rate for the mean fitness if $\approx 1.2$. Compering the two examples above, we see that the impact of the fitness landscape adaptation is  bigger than mutation matrix alone.

\subsection{Numerical example 3}
\begin{figure}[h!]
	\centering
	\includegraphics[width = 0.6\textwidth]{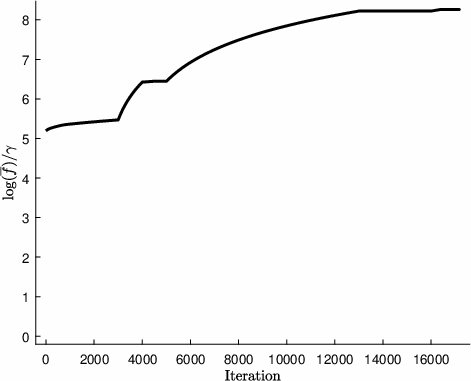}
	\caption{Example 3(subsequent $\bfM$ and $\bfQ$ changes): The mean fitness value changing over evolutionary time, which is represented by the number of iterations}
	\label{fig6}
\end{figure}

Let us assume that adaptation of the fitness landscape parameters and mutation matrix are two sequential processes. For this settings, we apply 3000 evolutionary steps for  mutation matrix change and 1000 for fitness landscape adaptation. 
\begin{itemize}
	\item The initial distribution of fitness: $\bfm_0: m_{01} =m_{0i} =0.0625, i=2,\ldots, 16$.  The death rates are the same as in the previous examples. 
	\item The first steady-state: $\bar{ u}(0) =$(0.0001, 0.0209, 0.0209, 0.1358, 0.0209, 0.1358, 0.1358, 0.5788, 0.0209, 0.1358, 0.1358, 0.5788, 0.1358, 0.5788, 0.5788, 1.9803) with the total population size $S = 5.194.$
	\item The step for evolutionary timescale iteration is $\varepsilon = 0.0001$.
\end{itemize}
Here, the fitness landscape transformation happens in 17172 iterations of the evolutionary  process. The mean fitness value dynamics is depicted in Fig.~\ref{fig6}. The distribution of  types ends up with $m_16=1, m_{i\neq 2} =0$ , $\bar{u}_{end}=$ (0.0022, 0.2217, 0.2217, 0.4375, 0.2217, 0.4375, 0.4375, 0.6493, 0.2217, 0.4375, 0.4375, 0.6493, 0.4375, 0.6493, 0.6493, 2.1501), and  $S(\bar{ u}_{end}) = 8.2621$.	In this case, the mean fitness increased by almost 1.6 times. 

The dynamics of the fitness landscape components is shown in Fig.~\ref{fig2}. In Fig.~\ref{fig3}, we show how the rates of 1st and 3d species change over this process: the fraction of the first one, dominant in the beginning, drops down, while the latter prevails.  
\begin{figure}[h!]
	\begin{center}
		\includegraphics[width = 0.6\textwidth]{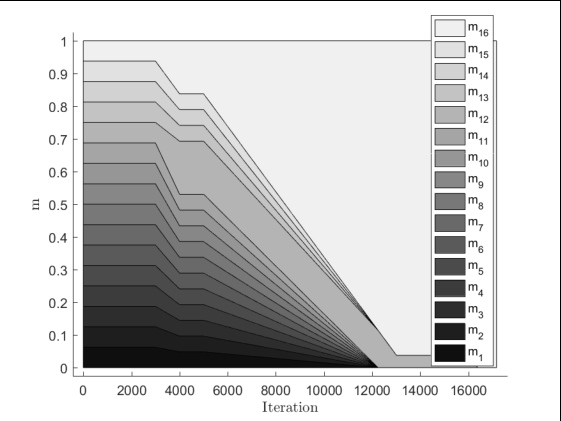}
		\caption{Example 3(subsequent $\bfM$ and $\bfQ$ changes): Dynamics of the fitness landscape parameters over evolutionary time: fitness matrix values $m_i$ in steady-states with respect to the number iterations}
		\label{fig7}
	\end{center}
\end{figure}

\begin{figure}[h!]
	\center{\includegraphics[width=0.6\linewidth]{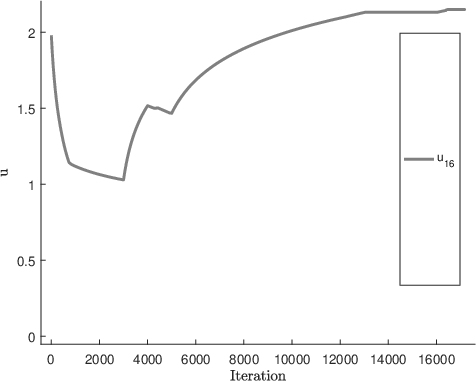} }
	\caption{Example 3(subsequent $\bfM$ and $\bfQ$ changes): Dependence of the steady-state distribution component $\bar{u}_{16}$ on the iteration number in the evolutionary timescale}
	\label{fig8}
\end{figure}

\begin{figure}[h!]
	\begin{minipage}[h]{0.5\linewidth}
		\center{\includegraphics[width=1\linewidth]{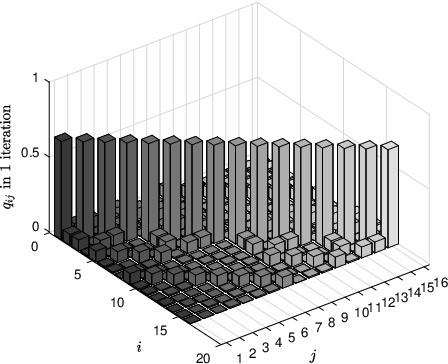} \\ a)}
	\end{minipage}
	\hfill
	\begin{minipage}[h]{0.5\linewidth}
		\center{\includegraphics[width=1\linewidth]{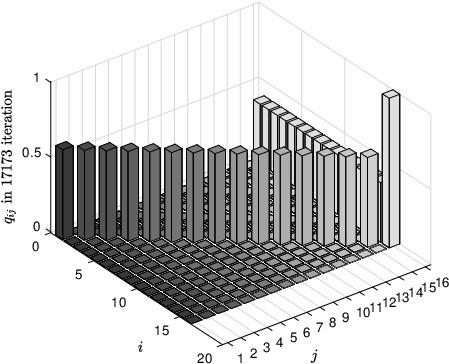}\\ b)}
	\end{minipage}
	\caption{Example 3(subsequent $\bfM$ and $\bfQ$ changes): The mutation matrix coefficients at a) the first and b)  17173d iterations of the evolutionary process}
	\label{fig9}
\end{figure}

\newpage5

\section{Conclusion}
In this paper, we applied an algorithm for the fitness landscape evolution of quasispecies systems. The main hypothesis and maximization technique were developed in the previous studies \cite{drozh18, drozh19} for general replicator systems and hypercycles. For these classes of systems, Fisher's theorem of natural selection gains a new mathematical interpretation. Taking into account two different timescales on which evolutionary process act, we formalized the extremum principle.  This approach allowed us to rewrite the initial problem in a form of sequential linear programming problems over corresponding steady-states. The numerical simulations showed that the mean fitness grow continuously over evolutionary time. At the same time, population structure and system parameters undergo significant changes, exploring evolutionary variability of the system.

\appendix
\section{Proof of Theorem}\label{app1}
Let $\bfu(t)$ be a solution to (\ref{eq12}) over the set, defined by (\ref{eq13}). Hence,
\begin{equation}
\bfD^{-1}\bfQ_m\bar{ \bfu} = e^{\gamma\bar{ S}}\bar{ \bfu} = \eta\bar{ \bfu}.
\end{equation}
Consider an adjoint problem:
\begin{equation}
	\left(\bfD^{-1}\bfQ_m\right)^T\bar{\bfv} = \lambda^*\bar{\bfv}.
\end{equation} 
Without loss of generality, we assume $\left(\bar{ \bfv}, \bar{ \bfu}\right) = 1.$
Then, 
\begin{equation}
	\eta = \left(\bfD^{-1}\bfQ_m\bar{\bfu}, \bar{\bfv}\right) = \left(\bar{\bfu}, \left(\bfD^{-1}\bfQ_m\right)^T\bar{\bfv}\right) = \lambda^*.
\end{equation}
Vice versa, if $\lambda^* = \eta,$ then $\lambda^* = e^{\gamma \bar{ S}}$. Since 
$$
\bfD^{-1}\bfQ_m\bar{\bfu} = \lambda^*\bar{\bfu} = e^{\gamma\bar{ S}}\bar{ \bfu}, 
$$
the vector $\bar{ \bfu}\in \mathbb{R}^n_{+}$ is a solution to (\ref{eq7})

\section{Second-order corrections}\label{app2}
Assume that all the elements $\bfM$ and $\bfQ$ are twice differentiable functions of the evolutionary time $\tau.$ In this case, the equality (\ref{eq20}) can be supplemented by another one for $\delta^2$:
\begin{equation}\label{b1}
	\delta^2\bfQ_m\bar{ \bfu}(\tau) + 2\delta\bfQ_m(\tau)\delta\bar{\bfu} + \bfQ_m\delta^2\bar{ \bfu} = \delta^2\lambda^*\bar{\bfu} + 2\delta\lambda^* + \lambda^*\delta^2\bar{ \bfu},
\end{equation}
where 
$$
\delta^2\bfQ_m(\tau) = \delta^2\bfM + 2\delta\bfQ\delta\bfM + \bfQ\delta^2\bfM, 
$$
and $\delta^2\bfQ(\tau), \delta^2\bfM(\tau), \delta^2\bar{\bfu}(\tau)$ have the elements $\frac{1}{2}q''_{ij},$ $\frac{1}{2}m''_{ij},$ and $\frac{1}{2}u''_{i}$ correspondingly. 
Let $\delta\lambda^*(\tau) = \delta\bar{ f}(\tau) = 0$ for all such elements that (\ref{eq20})
holds true. In this case, we have:
\begin{equation}
	\left(\bfQ_m-\lambda^*\bfo\right)\delta\bar{ \bfu} = -\bfD^{-1}\left(\delta\bfQ\bfM +\bfQ\delta\bfM\right)\bar{ \bfu} = -\bfD^{-1}\delta\bfQ_m\bar{ \bfu}.
\end{equation} 
The latter equation, taking into account Fredholm alternative, has a solution $\delta\bar{ \bfu}$ if and only if  the right-side of the equation is orthogonal to the solution to (\ref{eq21}). This condition is satisfied:
\begin{equation}\label{b3}
\delta\bar{ \bfu} =  -\left(\delta\bfQ_m - \lambda^*\bfo\right)^{-1} \bfD^{-1}\delta\bfQ_m\bar{ \bfu},
\end{equation}
or $\delta\lambda^* = 0$ 
The solution to (\ref{b3}) belongs to the subspace of vectors $W = \left\{\mathbf{w}: (\mathbf{w}, \bar{\bfu}) = 0\right\}.$
The vector $\bar{ \bfu}$ is an eigenvector of (\ref{eq12}) corresponding to the dominant eigenvalue $\lambda^*.$ Multiplying (\ref{b1}) by $\bar{ \bfv}$, which is the solution to the adjoint problem (\ref{eq21}), and having $\delta\lambda^*(\tau) = 0,$ we get:
\begin{equation}\label{b4}
	\delta^2\lambda^*(\tau) = \left(\delta^2\bfQ_m\bar{ \bfu}, \bar{ \bfv}\right)+
	2\left(\delta\bfQ_m\delta\bar{\bfu}, \bar{ \bfv}\right).
\end{equation} 
Substituting (\ref{b3}) into (\ref{b4}), we derive:
\begin{equation}\label{b5}
\delta^2\lambda^*(\tau) = \delta^2\bar{ f}(\tau) = \left(\delta^2\bfQ_m\bar{ \bfu}, \bar{ \bfv}\right) - 
2\left(\delta\bfQ_m\left(\bfQ_m-\lambda^*\bfo\right)^{-1}\bfD^{-1}\delta\bfQ_m\bar{\bfu}, \bar{ \bfv}\right).
\end{equation}
The sufficient condition for extremum is $\delta^2\bar{ f}(\tau)<0$ over the set (\ref{eq17}).

\section*{Acknowledgedments}
The work is supported by the Russian Science Foundation under grant 19-11-00008.

T.Y. was supported in part by an appointment to the National Library of Medicine (NLM) National Center for Biotechnology Information (NCBI) Research Participation Program. This program is administered by the Oak Ridge Institute for Science and Education through an interagency agreement between the U.S. Department of Energy (DOE) and the National Library of Medicine
(NLM). ORISE is managed by ORAU under DOE contract number DE-SC0014664. All opinions expressed in this paper are the author's and do not necessarily reflect the policies and views of NLM, DOE, or ORAU/ORISE.

\end{document}